\documentclass[a4paper,12pt]{article}
\usepackage[utf8x]{inputenc}
\usepackage{graphicx}
\usepackage{amssymb,amsmath}
\usepackage{textcomp}
\def\pacs#1{\vspace{10pt} \hspace{0.33cm} \rm PACS numbers: #1 \par \vspace{10pt}}
\usepackage{ctable}
\usepackage{graphicx}
\usepackage{subfigure}
\usepackage{placeins}
\usepackage[hyperindex=true,breaklinks]{hyperref}
\usepackage[english]{babel}
\usepackage{subfigure}
\usepackage{booktabs}
\usepackage{multicol}

\begin{document}

\title{Non-extensivity of hadronic systems}
\author{L. Marques, E. Andrade-II and A. Deppman}
\date{Instituto de Física, Universidade de São Paulo - IFUSP, Rua do Matão, Travessa R 187, 05508-900 São Paulo-SP, Brazil}

\maketitle


\begin{abstract} 
{The predictions from a non-extensive self-consistent theory recently proposed are investigated. Transverse momentum ($p_T$) distribution for several hadrons obtained in $p+p$ collisions are analyzed to verify if there is evidence for a limiting effective temperature and a limiting entropic index. In addition, the hadron-mass spectrum proposed in that theory is confronted with available data.

It turns out that all $p_T$-distributions and the mass spectrum obtained in the theory are in good agreement with experiment with constant effective temperature and constant entropic index. The results confirm that the non-extensive statistics plays an important role in the description of the termodynamics of hadronic systems, and also that the self-consistent principle holds for energies as high as those achieved in the LHC. A discussion on the best $p_T$-distirbution formula for fitting experimental data is presented.}
\end{abstract}

The Hagedorn's bootstrap idea based on a self-consistency requirement for the thermodynamics of fireballs predicted a limiting temperature for hadronic systems and also gave formulas for transverse momentum ($p_T$) distributions of secondaries and for the hadron-mass spectrum~\cite{Hagedorn}.

Experiments with $\sqrt{s}>$10~GeV, however, have shown that the $p_T$-distribution formula fails to describe the data. An empirical formula was proposed~\cite{Bediaga} including non-extensive statistics and it results that the modified formula can fit all available data for $p_T$-distributions. Although many works have been done on the subject, the use of non-extensive statistics in hadronic physics remains rather controversial.

Recently a non-extensive version of the self-consistency principle was proposed~\cite{Deppman}, leading to new formulas for mass spectrum and for transverse momentum distribution. The last one is similar to that proposed in Ref.~\cite{Bediaga}. In addition, the theory predicts a  limiting effective temperature and a limiting entropic index for all hadronic systems. The limiting effective temperature was predicted also in a parton-gas model with Tsallis distribution~\cite{Biro_Peshier}. These results establish a much more restrictive test to evaluate if the non-extensive statistics plays any role in the hot hadronic systems produced in high energy collisions.

   \begin{table}[!h]
    \centering
    \caption{Set of experimental data for $p+p$ collisions.}
    \begin{tabular}{cccc}\toprule
    Experiment                             &    Particle              &	Reference     \\      \midrule
    ALICE (LHC)                               &  $\pi^{0}$, $\eta$ &  \cite{1Alice1} \\
    ALICE (LHC)                             &    $\phi$, $\omega$   & \cite{5Alice2} \\
    ALICE (LHC)                              &   $\pi^{\pm}$, $P^{\pm}$, $K^{\pm}$    & \cite{8Alice3} \\
    ATLAS (LHC)                               &    $J/\psi$  & \cite{4Atlas1} \\
    CMS (LHC)                             &     $J/\psi$  & \cite{9cms1} \\
    CMS (LHC)                             &     $\Lambda_{b}^{0}$  & \cite{11cms2} \\
    CMS (LHC)                             &     $K^{0}_{S}$, $\Lambda$, $\Xi^{-}$  & \cite{12cms3} \\
    LHCb (LHC)                               &    $B^{+}$ & \cite{3lhcb1} \\
    LHCb (LHC)                             &     $\phi$  & \cite{6lhcb2} \\
    LHCb (LHC)                             &     $J/\psi$  & \cite{10lhcb3} \\
    \bottomrule
    \end{tabular}
    \label{tab:Expe}
    \end{table}

In this work experimental data for $p_T$-distributions from different experiments and for several hadrons produced in $p+p$ collisions at ultrarelativistic energies are analysed in order to investigate the theoretical predictions given in Ref.~\cite{Deppman}. Also, the theoretical mass spectrum is compared to experimental data. The experimental data for $p_T$-distributuions used in the present analysis are summarized in Table~\ref{tab:Expe}.

Initially it is important to clarify that the $p_T$-distribution given by
\begin{equation}
 \frac{d^2N}{dp_T\,dy}=gV\frac{p_T m_T coshy}{(2\pi)^2}\bigg(1+(q-1)\frac{m_Tcoshy-\mu}{T} \bigg)^{-\frac{q}{q-1}}
\label{nonext}
\end{equation}
can be directly obtained from the Tsallis entropy~\cite{Tsallis1998} through the usual thermodynamical relations~\cite{Cleyman_Worku}. Despite this fact, in many analysis other $p_T$-distributions are used~\cite{1Alice1,5Alice2,8Alice3,11cms2}, as
\begin{equation}
 \frac{d^2N}{dp_T\,dy}=p_T\frac{dN}{dy}\frac{(n-1)(n-1)}{nC[nC+m_o(n-2)]}\bigg(1+\frac{m_T-m_o}{nC} \bigg)^{-n}\,.
\label{tail}
\end{equation}
In the equations above, $y$ is the hadron rapidity, $\mu$ is the chemical potential, $m_T=\sqrt{p_T^2+m_o^2}$, with $m_o$ being the hadron mass, $n$ and $C$ are constants, $V$ is the volume and $g$ is the degeneracy factor.

The $p_T$ dependence in both formulas can be made quite similar by adopting~\cite{Cleyman_Worku}
\begin{equation}
 n = \frac{q}{q-1}
\end{equation}
and
\begin{equation}
 nC = \frac{T}{q-1}\,,
\end{equation}
but the factor $m_T$ present in Eq.~\ref{nonext} and absent in Eq.~\ref{tail} is sufficient to produce very different values for the parameters $T$ and $q$ when those equations are used to fit experimental data, even if quite good fittings are obtained with both equations.

   \begin{figure}[!h]
       \centering
                   {\label{fig:T_H-T-p+p}\includegraphics[width=14.cm]{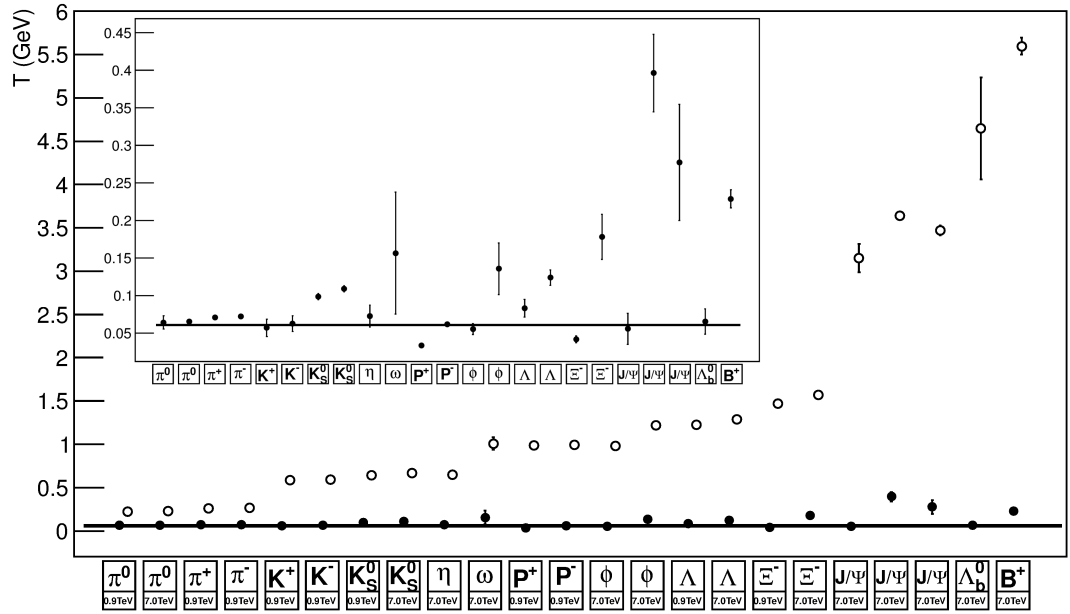}}
          \caption{Effective temperature, $T$, resulting from the fittings of Eq.~\ref{fittingformula} (full circles), assuming $y=0$ and $\mu=0$, and Eq.~\ref{tail} (open circles). The full square indicates the result obtained from the mass spectrum analysis (see text). The inset shows the effective temperature obtained through the use of Eq.~\ref{fittingformula} in more details. Full lines indicate the constant value, $T_o$, which best fits the data (full symbols).}
       \label{figT}
   \end{figure}

For the fittings of experimental data for $p_T$ distributions we follow Cleymans and Worku~\cite{Cleyman_Worku} and use $y=0$ and $\mu =0$ in Eq.~\ref{nonext}, resulting in
\begin{equation}
 \frac{d^2N}{dp_T\,dy}\bigg|_{y=0}=gV\frac{p_T m_T }{(2\pi)^2}\bigg(1+(q-1)\frac{m_T}{T} \bigg)^{-\frac{q}{q-1}}\,.
\label{fittingformula}
\end{equation}
Since experiments report data for  relatively small ranges of rapidity this can be considered an appropriate approximation.

In Fig.~\ref{figT} the effective temperatures obtained from those fittings are presented. It is clear that the temperature obtained with Eq.~\ref{tail} varies in a broad range, systematically increasing with the hadron mass. The results obtained with Eq.~\ref{fittingformula}, on the other hand, give temperatures spread over a much narrower range around a constant value $T_o=(60.7 \pm 0.5)$~MeV (full lines in Fig.~\ref{figT}).

Comparing Eqs~\ref{fittingformula} and~\ref{tail}, it is easy to understand that the absence of the $m_T$ factor in the latter gives rise to the increasing temperature behaviour observed in Fig.~\ref{figT}. Indeed, the effects of the increase in $m_T$ due to the increase of $p_T$ in Eq.~\ref{fittingformula} are reproduced in Eq.~\ref{tail} by an increase of $T$.

\begin{figure}[!h]
       \centering
                   {\label{fig:T_H-T-p+p}\includegraphics[width=14.cm]{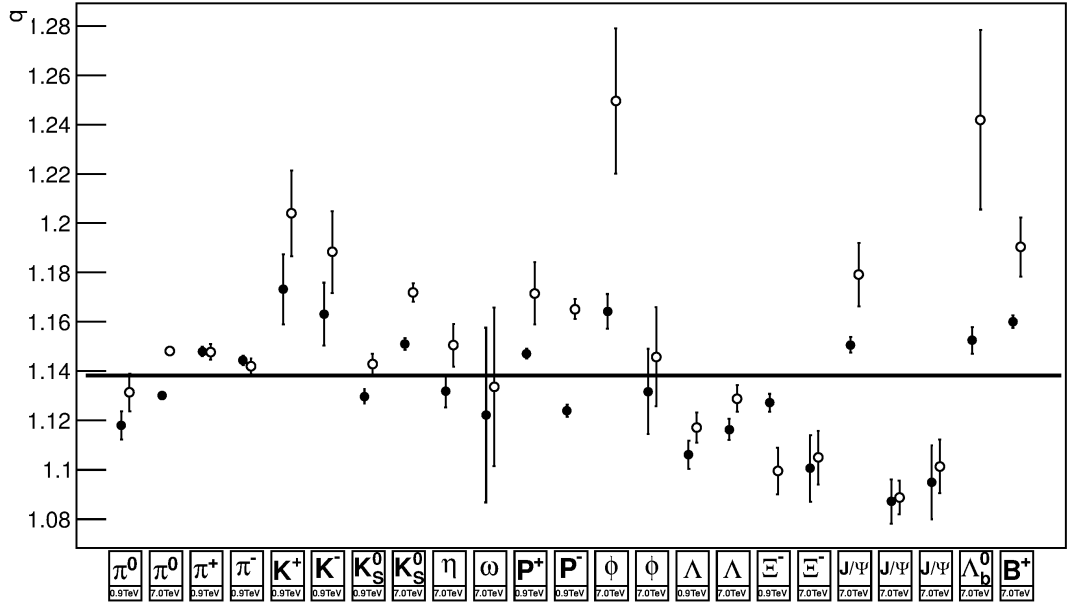}}
          \caption{Entropic factor, $q$, resulting from the fittings of Eq.~\ref{fittingformula} (full circles) and Eq.~\ref{tail} (open circles). The full square indicates the result obtained from the mass spectrum analysis (see text). The full line indicates a constant $q$ fitted to the data obtained with Eq.~\ref{fittingformula}.}
       \label{figq}
   \end{figure}

The results of the entropic index obtained from the fittings are plotted in Fig.~\ref{figq}. In this case, for both equations the results are spread around an average value with no evidence of a systematic trend. But again the values are spread over a broader range in the case of Eq.~\ref{tail}, while they are limited to a narrower range around $q_o=1.138 \pm 0.006$ when Eq.~\ref{fittingformula} is used.

In principle, there is nothing wrong with an increasing temperature for increasing hadron mass. The relevant points here are:
\begin{itemize}
 \item[i)] The $p_T$-distribution obtained from Tsallis entropy by using the usual thermodynamical relations is Eq.~\ref{fittingformula}, not Eq.~\ref{tail}.
\item[ii)] If Tsallis statistics is the basis for a thermodynamical description of hadronic systems, then the self-consistency principle lead to a limiting effective temperature. Such limiting temperature is observed when Eq.~\ref{fittingformula} is used, as shown by the results in Fig.~\ref{figT}.
\item[iii)] The theory predicts a limiting entropic index, which is also observed in the analysis of $p_T$-distribution, as shown in Fig.~\ref{figq}.
\end{itemize}

These results are in agreement with recent analysis performed in Refs.~\cite{Cleyman_Worku,Sena, Sena1}, where constant temperature and entropic index were found with values similar to those obtained here. The present analysis extends those analysis by considering identified particles and by including $p+p$ collisions up to $\sqrt{s}$=7~TeV.
Regarding the parameter $V$ in Eq.~\ref{fittingformula}, which is also a free parameter, we cannot perform a systematic study of its values since we use in this analysis experimental yields obtained in different experiments and different laboratories.

 The results discussed above show that there are strong evidences that the non-extensive statistics plays an important role in the thermodynamical description of hadronic systems and that the self-consistency conditions find support in the experimental data from ultrarelativistic collisions. It is worthwhile to stress the importance of using the $p_T$-distribution formula which is consistently derived from Tsallis entropy by the use of thermodynamical relations.

A crucial verification of the theory is related to the mass spectrum. In fact, if Hagedorn's theory fails to describe $p_T$-distributions for $\sqrt{s}>$10~GeV, it also has problems to describe the data for hadron-mass spectrum. The Hagedorn temperature, $T_H$, varies from 141~MeV up to 340~MeV, depending on the parametrization used for the mass spectrum formula, specially for the multiplying factor\cite{CWmass,Chojnacki1,Chojnacki2,Chatterjee,Broniowski1,Broniowski2,Megias1,Megias2,Arriola}. For the most used parametrization, however, $T_H$ is much higher than that expected from hadron-hadron collisions, where $T_H \approx$160~MeV~\cite{CWmass,Chatterjee}.

According to the non-extensive self-consistent theory, the hadron-mass spectrum is given by
\begin{eqnarray}
\rho(m) = \gamma m^{-5/2} e_q^{\beta_o m}\,,
\label{mass_spectrum}
\end{eqnarray}
where $e_q^{x}$ is the q-exponencial function~\cite{Deppman} given by
\begin{equation}
 e_q^x=[1+(q-1)x]^{1/(q-1)}\,.
\end{equation}
It is important, therefore, to verify if this equation can describe the mass spectrum data with the same values $T_o$ and $q_o$ obtained in $p+p$ collisisons. A power-law mass spectrum was already used in Ref.~\cite{Biro_Peshier}.

The cumulative hadron-mass distribution is given by
\begin{equation}
 r(m)=\int \rho(m) dm=\frac{-2\gamma}{3} m^{-3/2} \,_2F_1\bigg(-\frac{3}{2},-\frac{1}{q-1};-\frac{1}{2};-(q-1)\beta m \bigg)+k\,,
\label{cumulative_spectrum}
\end{equation}
where $k$ is a constant and $_2F_1(a,b;c;z)$ is the Gauss' hypergeometric function. This equation was fitted to the available data for cumulative mass spectrum~\cite{CWmass}.

\begin{figure}[!h]
        \centering
                    {\label{fig:T_H-T-p+p}\includegraphics[width=14.cm]{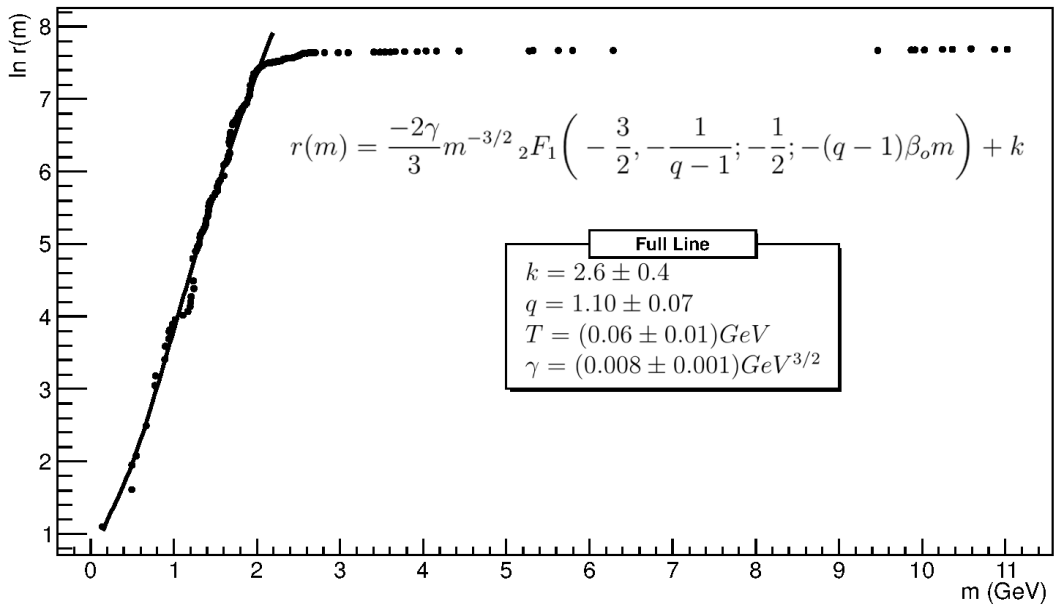}}
           \caption{Cumulative hadron-mass spectrum. Full line represents the calculation with Eq.~\ref{cumulative_spectrum} using $k=$2.7$\pm$0.4, $\gamma=$(5$\pm$3)10$^{-3}$ GeV$^{3/2}$, $q=$1.103$\pm$0.007 and $T=$(52$\pm$7) GeV. Full-circles represent the available data taken from Ref.~\cite{CWmass}.}
      \label{figMS}
    \end{figure}

In Fig.~\ref{figMS} the best fitted curve is shown, and it is possible to observe a good agreement between data and calculation. The fitting procedure does not take into account data above 2~GeV, since the information above this threshold is not considered reliable. This procedure is usual in the study of mass spectrum~\cite{CWmass,Chatterjee}.

The curve in Fig.~\ref{figMS} is obtained from Eq.~\ref{cumulative_spectrum} with $T=($52$\pm$7)~MeV and $q=$1.103$\pm$0.007. These values fall in the same range of the corresponding ones obtained in $p+p$ analysis, as shown in Figs~\ref{figT} and~\ref{figq}. Therefore a good agreement is found between the results from $p_T$-distribution analysis and from mass spectrum analysis. It is possible to conclude that the non-extensive self-consistent theory proposed in Ref.~\cite{Deppman} can describe simultaneously the $p_T$-distribution and the hadron-mass spectrum with constant effective temperature and constant entropic index.

A final remark can be made about the values $T_o$ and $q_o$. According to the interpretation of non-extensivity given in Ref.~\cite{Wilk2009a}, $T_o$ and $q_o$ are related to the critical temperature by
\begin{equation}
 T_o=T_H+(q_o-1)\,c\,,
\label{Tqrelat}
\end{equation}
where $c$ is a constant depending on the energy transfer between the source and its surroundings and on thermodynamical properties of the medium~\cite{Wilk2009a, Wilk2009b}. In Refs.~\cite{Sena,Sena1} it was shown that $T_H=($192$\pm$15)~MeV and $c=$-(950$\pm$10)~MeV. It is interesting to observe that $T_H$ is in good agreement with the critical temperature from lattice QCD~\cite{Ejiri,Miura}. It is also clear that the values found here for $T_o$ and $q_o$ satisfy Eq.~\ref{Tqrelat}.

In conclusion, this work presents an extensive analysis of $p_T$-distribution from $p+p$ collisions at ultrarelativistic energies in order to test the predictions of the non-extensive self-consistent theory proposed in Ref.~\cite{Deppman}. The results show a limiting effective temperature $T_o=(60.7 \pm 0.5)$~MeV and a limiting entropic index $q_o=1.138 \pm 0.006$.

Also the theoretical mass spectrum is compared with the available data resulting in good agreement between calculation and data for $T=($52$\pm$7$)$~MeV and $q=$1.103$\pm$0.007. These values are within the range of the values $T_o$ and $q_o$ found in $p_T$-distribution analysis.

Finally, the relation between $T_o$, $q_o$ and  $T_H$ shows that the results obtained in the present work are in agreement with the critical temperature predicted in lattice QCD calculations.

With these results it is possible to observe that the non-extensive self-consistent thermodynamical approach can describe the main features of the hadronic system formed in high energy collisions. We do not claim that it describes all possible aspects of the problem, since this thermodynamical theory deals only with a system in its stationary state and therefore it is not supposed to explain what happens before this state is reached nor what happens after the freeze-out.

The results obtained here allow a complete thermodynamical description of dilute hadronic systems. With this description one can, for instance, compares the predictions of the non-extensive self-consistent theory with lattice-QCD results~\cite{Deppman2}. In addition, this work indicates that microscopical mechanisms that lead to non-extensivity should be included in resonance hadron gas models. Some hints on how to pursue this objective can be found in Refs.~\cite{Biro_Peshier, Wilk2009a, Borland}.

\section{Acknowledgements}
This work received support from the Brazilian agency, CNPq, under grant 305639/2010-2 (A.D.), and by FAPESP under grant 2012/05085-1 (LM).

\bibliographystyle{aipproc}

\end{document}